\documentclass{iauc}
\usepackage{graphicx}
\pubyear{2005}
\volume{199}
\pagerange{1--}
\jname{Probing Galaxies through Quasar Absorption Lines}
\editors{P. R. Williams, C. Shu, and B. M\'{e}nard, eds.}

\title[DLA Evolution] 
{The Evolution of Damped Ly-$\alpha$ Absorbers: Metallicities and Star 
Formation~Rates}

\author[V. P. Kulkarni et al.]   
{Varsha P. Kulkarni$^1$,
Donald G. York$^2$, James T. Lauroesch$^3$,
 S.~Michael Fall$^4$, Pushpa Khare$^5$, Bruce E. Woodgate$^6$, Povilas 
 Palunas$^7$, Joseph Meiring$^1$, Deepashri G. Thatte$^1$, Daniel E. Welty$^2$, 
 \break \and James W. Truran$^2$}

\affiliation{$^1$Dept. of Physics and Astronomy, Univ. of South Carolina,
Columbia, SC 29208, USA \break 
emails: kulkarni@sc.edu; meiring@physics.sc.edu; thatte@physics.sc.edu\\[\affilskip]
$^2$Dept. of Astronomy and Astrophysics, Univ. of Chicago, 
Chicago, IL 60637, U.S.A. \break 
email:don@oddjob.uchicago.edu\\[\affilskip]
$^3$Dept. of Physics and Astronomy, Northwestern Univ.,
Evanston, IL 60208, USA \break 
email: jtl@elvis.astro.northwestern.edu\\[\affilskip]
$^4$Space Telescope Science Inst., Baltimore, MD 21218,  U.S.A. \break 
email:fall@stsci.edu\\[\affilskip]
$^5$Dept. of Physics, Utkal University, Bhubaneswar, 751004, India 
\break 
email:khare@iopb.res.in\\[\affilskip]
$^6$NASA/Goddard Space Flight Center, Code 681, 
Greenbelt, MD 20771, U.S.A.  \break 
email:woodgate@uit.gsfc.nasa.gov\\[\affilskip]
$^7$McDonald Observatory, Univ. of Texas, \break Austin, TX 78712, U.S.A.  
\break 
email: palunas@astro.as.utexas.edu}

\begin{document}

\maketitle

\begin{abstract}
The damped Lyman-$\alpha$ 
(DLA) and sub-DLA quasar absorption lines  
provide powerful probes of the evolution of metals, gas, and stars in 
galaxies. One major obstacle in trying to understand the evolution 
of DLAs and sub-DLAs has been 
the small number of metallicity measurements at $z < 1.5$, an epoch 
spanning $\sim 70 \%$ of the cosmic history. In recent surveys 
with the Hubble Space Telescope and Multiple Mirror Telescope, we have 
doubled the DLA Zn sample at $ z < 1.5$. Combining our results with 
those at higher redshifts from the literature, we find that 
the global mean metallicity of DLAs does not rise to the solar value at low 
redshifts.   
These surprising results appear to contradict the near-solar mean metallicity 
observed for nearby ($z \approx 0$) galaxies and the predictions of 
cosmic chemical evolution models based on the global star formation history. 
Finally, we discuss direct constraints on the star formation rates (SFRs) in the  
absorber galaxies from our deep Fabry-Perot Ly-$\alpha$ imaging study and  
other emission-line studies in the literature.  
A large fraction of the observed heavy-element quasar absorbers at $0 < z < 3.4$
appear to have SFRs substantially below the global mean SFR, consistent with  
the low metallicities observed in the spectroscopic studies. 

\end{abstract}

\firstsection 
\section{Introduction}

Heavy-element quasar absorption systems probe galaxies at various epochs, 
selected independent of their luminosities. The damped Ly-alpha absorbers 
(DLAs; log $N_{\rm H I} > 20.3$) and sub-DLAs 
(19.0 $<$ log $N_{\rm H I} < 20.3$) constitute a large fraction of H I 
in galaxies (e.g. Wolfe et al. 1995; Peroux et al. 2003), and provide the best existing 
probes of the chemical composition 
of galaxies over $\sim 90 \%$ of the cosmic history. They should thus provide important 
clues to the history of metal production 
and star formation in galaxies. 

\section{Evolution of Metallicity}

Most cosmic chemical evolution models predict the global mean interstellar 
metallicity of galaxies to rise with time, from low values at high redshifts 
to solar or near-solar values at $z=0$ (e.g., Pei \& Fall 1995; Malaney 
\& Chaboyer 1996; Pei, Fall, \& Hauser 1999; Somerville et al. 2001). This rise 
in metallicity at low redshifts is driven by the high global star formation 
rate (SFR) at $z \gtrsim 1.5$ implied by galaxy imaging surveys such 
as the  Canada-France Redshift Survey (CFRS) and the Hubble Deep Field (HDF). 
(see, e.g., Lilly et al. 1996; Madau et al. 1996).  
Indeed, 
the mass-weighted mean metallicity of nearby galaxies is also near-solar 
(e.g. Kulkarni \& Fall 2002). Do the DLA data show the predicted rise in the 
global mean metallicity with decreasing redshift? 

We adopt Zn as our primary metallicity indicator because (a) 
Zn tracks Fe closely, to within 
$\sim \pm 0.1$ dex, in Galactic halo and disk stars 
for [Fe/H] $\gtrsim -3$; (b) Zn is relatively 
undepleted on interstellar dust grains; and (c) the Zn II 
$\lambda \lambda 2026, 2062$ lines in DLAs 
often lie outside the Lyman-$\alpha$ forest and are usually unsaturated. 
In the absence of selection effects, the quantity 
$\Omega_{\rm{metals}}^{\rm ISM}/\Omega_{\rm{gas}}^{\rm ISM}$ 
is equal to the   
$N_{\rm H I}$-weighted mean metallicity $\overline{Z}$ in a sample 
of DLAs, where 
$\overline{Z} / Z_{\odot} = 
[{\Sigma N({\rm Zn \, II})_{i} / \Sigma N({\rm H \, I})_{i} ] /
{({\rm Zn/H})_{\odot}}}$ (e.g., Lanzetta et al. 1995; Kulkarni \& Fall 2002).    
There has been considerable debate about whether or not this quantity 
$\overline{Z}$ rises with decreasing redshift. 
Based on 57 Zn measurements 
at $0.4 < z < 3.4$, Kulkarni \& Fall (2002) 
found the slope of the 
metallicity-redshift relation to be $-0.26 \pm 0.10$, consistent 
at $\approx 2-3 \, \sigma$ level with both the predicted rates of evolution 
(-0.25 to -0.61), and with no evolution. Prochaska et al. (2003) 
reached similar conclusions (slope $-0.25 \pm 0.07$) for $0.5 < z < 4.7$, 
combining Zn, Fe, Si, S, O and X-ray absorption measurements in 121 DLAs.  
The main reason for the debate about this issue 
is the small number of measurements available, especially at $z < 1.5$. 
Space-based ultraviolet (UV) measurements are needed  
to access the H~I Ly-$\alpha$ lines at $z < 1.6$ and the Zn II lines at 
$z < 0.6$. Furthermore, for $0.6 < z < 1.5$, the Zn II lines lie in the 
blue region, where most spectrographs have relatively low sensitivity. 
To improve this situation, we have recently started to expand the DLA abundance 
samples at $z < 1.5$, using the Multiple Mirror Telescope 
(MMT) and the Hubble Space Telescope (HST).  

\subsection{Abundance Measurements for DLAs at $z < 1.5$}

The low-$z$ end is important to clarify the overall shape of the 
metallicity-redshift relation and to understand the relation of DLAs to 
present-day galaxies. To constrain the low-$z$ end better, we carried out 
HST STIS observations of 4 DLAs with $0.09 < z < 0.52$ and 
$20.3 <$ log $N_{\rm H I} < 21.3$ (Kulkarni et al. 2005a). 5-10 orbits were 
spent per object with the STIS G230M/NUV-MAMA or G230MB/CCD configurations 
at spectral resolutions of 10,400-14,300. The data were reduced  
using IRAF and STSDAS/CALSTIS package. Column densities were estimated by  
fitting the line profiles and verified using the apparent optical depth method. All three of 
the four DLAs in our sample, where we could put meaningful constraints on the 
Zn abundance, have Zn/H below 10-20 $\%$ 
solar. 

To expand the DLA Zn samples in the intermediate-redshift range, we have been 
carrying out a spectroscopic survey of DLAs with $0.6 < z < 1.5$ using the MMT 
blue channel spectrograph. Despite the modest resolution ($\sim 75$ km s$^{-1}$) 
of this instrument, we have been able to detect lines of Zn, Cr, Fe, Mn, Ni, Ti 
etc. because of the high S/N in our spectra. Some of the data from this ongoing 
survey are described by Khare et al. (2004). The Zn abundances for the DLAs 
in this sample were found to be 3-32 $\%$ solar. 

\subsection{Implications for the Global Metallicity-Redshift Relation}
Our HST and MMT data have so far doubled the DLA Zn sample at $z < 1.5$ and 
tripled the sample at $z < 1$. Fig. 1 shows the logarithm of the $N_{\rm H I}$-weighted mean Zn metallicity  
as a function of redshift, for the sample of 
87 DLAs, based on our HST and MMT 
results so far and the values available in the literature 
(see Kulkarni et al. 2005a  
and references therein). 
Horizontal bars for each bin show the range of 
DLA redshifts within that bin. The vertical error bars 
show the 1 $\sigma$ uncertainties in the $N_{\rm H I}$-weighted mean 
metallicities, and include both the measurement uncertainty and the sampling 
uncertainty. The left panel shows the ``maximum-limits'' case, where the Zn upper limits 
are treated as detections. The filled circles  
in the right panel show the 
``minimum-limits'' case with the Zn limits treated as zeros. The unfilled 
circles in the right panel show the 
modified minimum-limits sample, using other elements to constrain 
the metallicities in cases of Zn limits.  The slopes of the metallicity-redshift 
relation in the range $0.09 < z < 3.90$ for these three cases are   
$ -0.18 \pm 0.06$, $-0.22 \pm 0.08$, and $ -0.23 \pm 0.06$ . 
The corresponding estimates of the $z = 0$ intercept of the metallicity-redshift 
relation are $-0.74 \pm 0.15$, $-0.75 \pm 0.18$, and $-0.71 \pm 0.13$, 
respectively. Clearly, the global mean metallicity of DLAs does not seem to 
rise up to the solar value at low redshifts, and shows at best a slow evolution 
at a rate of $\approx 0.2$ dex or less per unit redshift. 
This result appears to contradict the predictions from most cosmic chemical 
evolution models and the global star formation history. We return to the implications 
of this in section 4. 

\begin{figure}
 \includegraphics[height=2in,width=5in,angle=0]{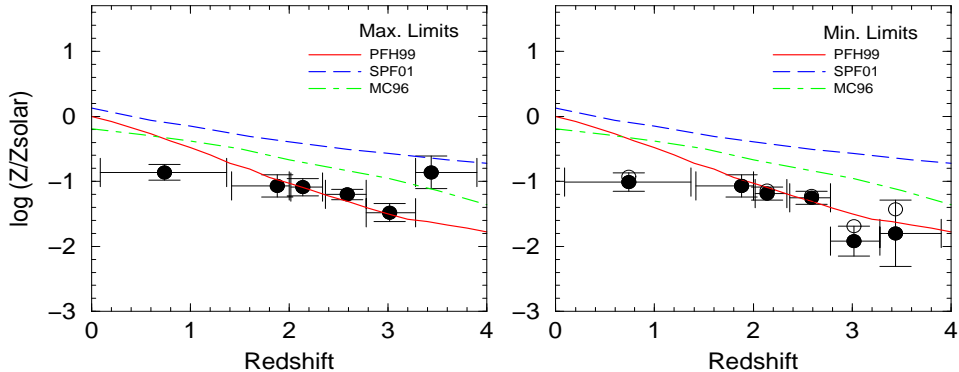}
  \caption{Global metallicity-redshift relation for 
DLAs from 87 measurements, based on our HST and MMT data  
and the literature. The circles show the logarithm of the  
$N({\rm H \, I})$-weighted mean Zn metallicity relative to the solar value. 
The filled circles in the left and right panels 
show the results with the Zn upper limits treated as detections and zeros, 
respectively.  The unfilled 
circles in the right panel show the results using information from other 
elements in those cases where only limits rather than detections are 
available for Zn.  
The solid, short-dashed, and dot-dashed curves show, 
respectively, the ``true'' mean metallicity (not corrected for dust 
obscuration) expected in the cosmic chemical evolution models 
of  Pei et al. (1999), Somerville et al. (2001), and 
Malaney \& Chaboyer (1996).}\label{fig:Zvsz_DLA}
\end{figure}

\section{Evolution of Star Formation Rates}

Another important quantity necessary to study the evolution of an absorber 
galaxy is its SFR. The SFR 
can be estimated from emission lines such as Ly-$\alpha$, H-$\alpha$, 
[O II] or [O~III] seen commonly in star-forming regions. 
Searches for low-$z$ DLAs have often imaged, 
and sometimes spectroscopically 
confirmed the absorbing galaxies (e.g., Yanny et al. 1990; 
LeBrun et al. 1997; Chen et al. 2005). However, this has been much more difficult at high redshifts. 
There have been a few detections of Ly-$\alpha$ emission in high-$z$ quasar 
absorber fields (e.g. Lowenthal et al. 1991; Francis et al. 1996; 
Roche et al. 2000). However, 
most other attempts to detect Ly-$\alpha$ emission from high-$z$  
intervening ($z_{abs} < z_{em}$) DLAs have produced either 
non-detections or weak detections. 
Many of the few confirmed Ly-$\alpha$ 
detections for high-redshift DLAs have been 
for absorbers with $z_{abs} \approx z_{em}$, which may differ from the 
cosmologically more interesting general population of DLAs with 
$z_{abs} < z_{em}$. Several attempts to 
detect DLAs in H-$\alpha$ have either yielded non-detections or detected 
companions 
separated by large angular distances from the quasars, rather than 
objects close to quasar sightlines. 

We have been carrying out a narrow-band Fabry-Perot (FP) imaging survey 
of quasar absorber fields at the Apache Point Observatory 3.5 meter telescope 
(Kulkarni et al. 2005b). 
The blue and vis-broad etalons in the Goddard Space Flight Center (GSFC) Fabry 
Perot (FP) imaging system have optimimum sensitivity and resolution in the 
wavelength range $\sim 4000-5000$ {\AA}. We 
therefore restricted our search to the redshift range $2.3 < z < 3.1$. 
We searched the York et al. (1991) catalog of heavy-element quasar absorbers 
for absorbers with (i) $2.3 < z_{abs} < 3.1$  (ii) $z_{abs} < z_{em} - 0.6$ 
to avoid absorbers possibly associated with the quasars, and (iii) with 
well-detected mixed-ionization lines (Si II, Al II or O I in addition to C IV and/or Si IV). Six 
such absorbers, including 2 DLAs, were finally observed. 

The observations were carried out during 9 runs between October 2000 and 
May 2004 at the 3.5 meter Apache Point Observatory (APO) telescope. Total integration 
times were 320-600 minutes per field, making these among the deepest images ever taken 
for quasar absorber fields. Data reduction was carried out using standard IRAF 
tasks. Figures 2a and 2b show the reduced broad-band and narrow-band images of 
one of our fields, Q0216+080, which is known to have a DLA at $z=2.2931$. 
The stripes at the borders of some images are an artifact of the coadding of 
the dithered images. The quasar is considerably dim in the 
narrow-band image because of the presence of the foreground DLA. 
 Fig. 2c shows the continuum-subtracted image, obtained by aligning the images 
spatially, 
and adjusting the scaling factor so as to minimize variance in the central 
portion of the subtracted image. 
All of the objects in the narrow-band images disappeared almost completely 
after subtraction of the continuum. The slight residuals left 
at the positions of some stars are because of the difficulty in 
matching the point spread functions (PSFs) perfectly in the broad-band and narrow-band images. 
No significant Ly-$\alpha$ emission was detected from any 
object in this or any of the other quasar absorber fields in our sample. 
To estimate the limits on the Ly-$\alpha$ fluxes for 
the absorbers in these fields, we used our observations of the 
field of the radio galaxy 53w002 with known Ly-$\alpha$ 
emitters at $z = 2.39$ (Pascarelle et al. 1996), 
and observations of a standard star. 
Based on this, the 3 $\sigma$ observed-frame Ly-$\alpha$ flux sensitivity reached 
in our images is in the range of $1.9-5.4 \times 10^{-17}$ 
erg s$^{-1}$ cm$^{-2}$, implying 
3 $\sigma$ SFR 
limits of 0.8-2.4 M$_{\odot}$ yr$^{-1}$.

\begin{figure}
 \includegraphics[height=1.7in,width=1.7in,angle=0]{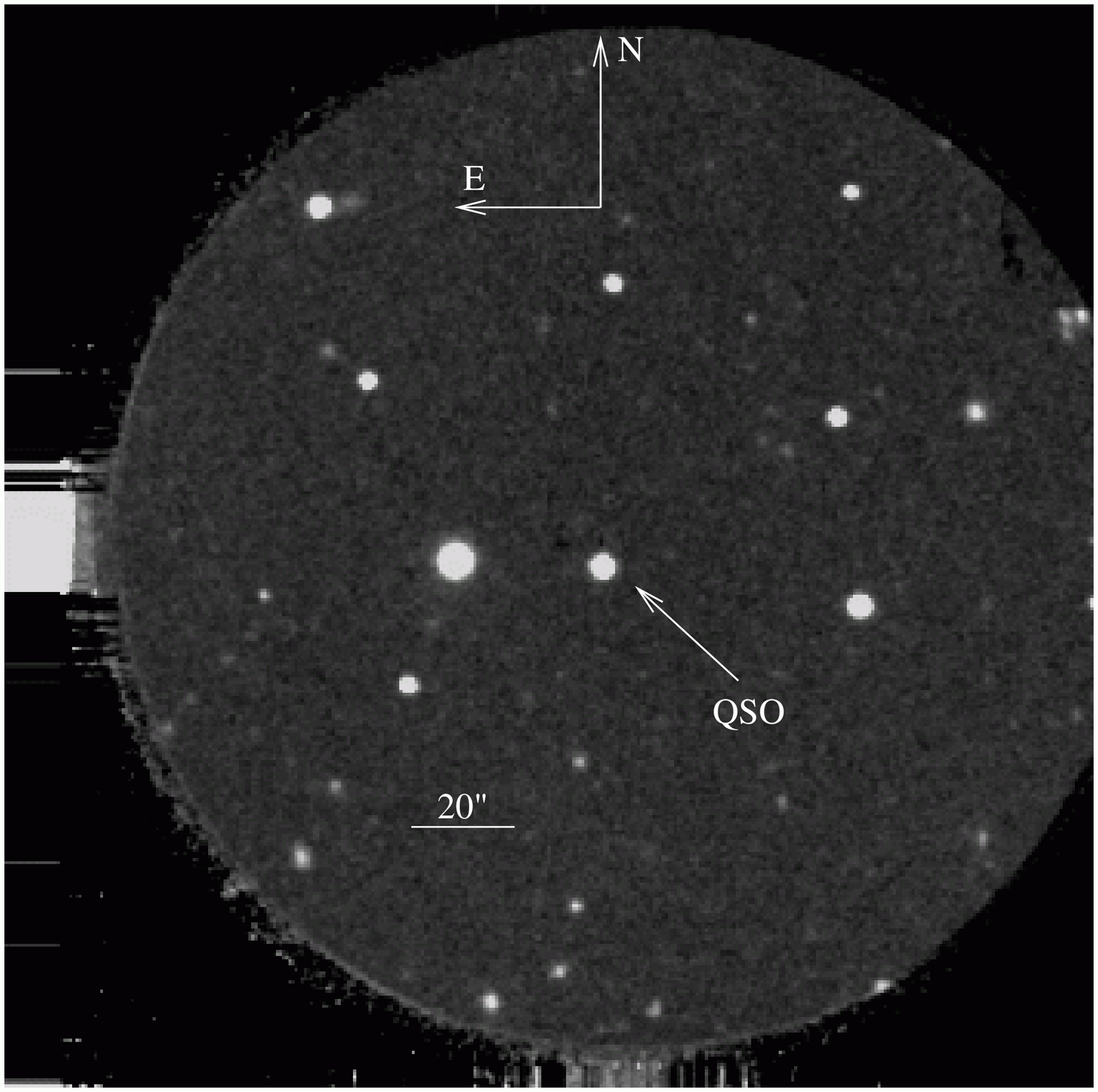} \hskip 5pt
 \includegraphics[height=1.7in,width=1.7in,angle=0]{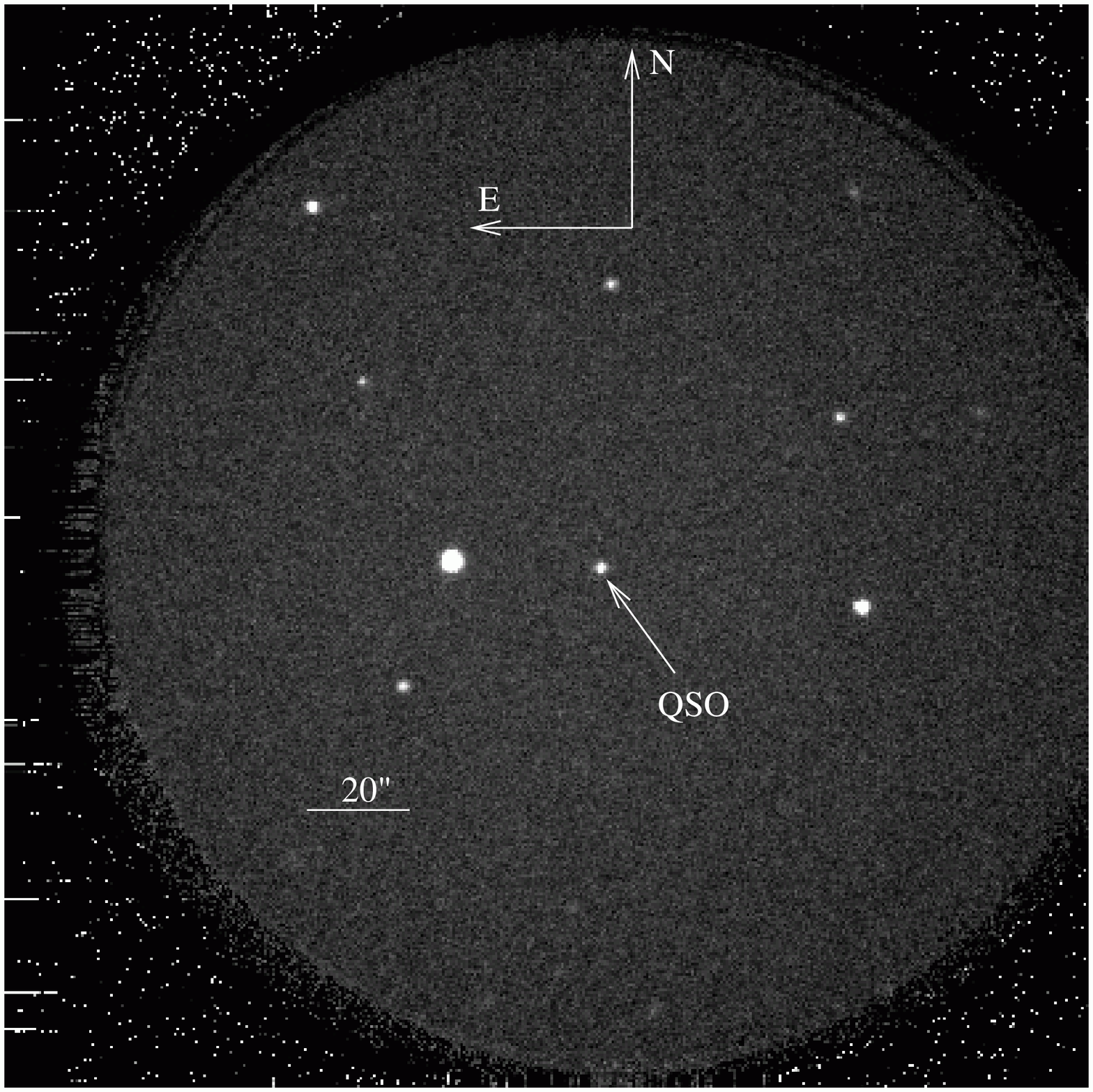} \hskip 5pt
 \includegraphics[height=1.7in,width=1.7in,angle=0]{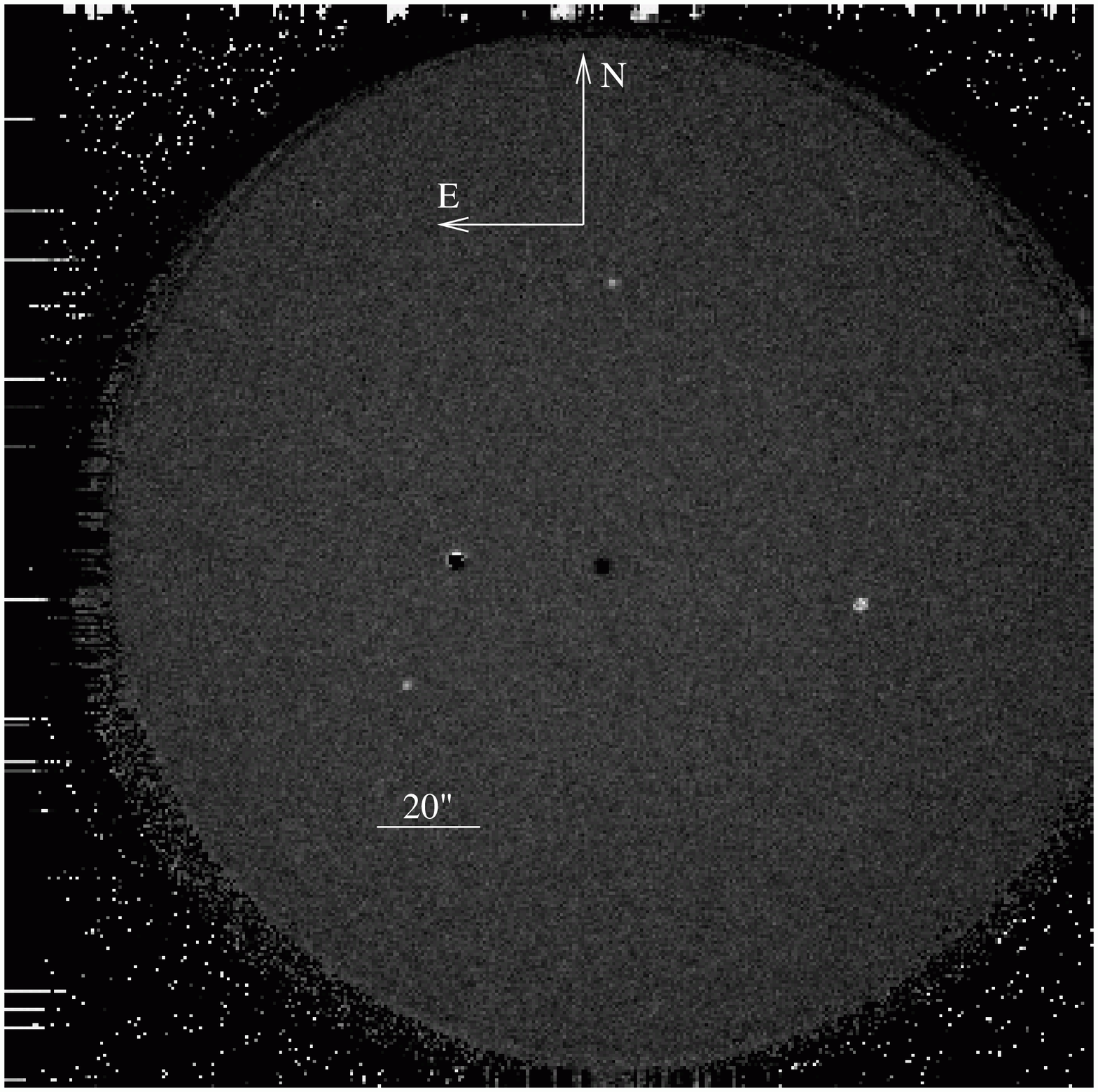}
  \caption{APO images of the field of Q0216+080  
in B band (left) and in narrow-band centered on Ly-$\alpha$ 
emission at 
$z = 2.2931$ before (center) and after (right) continuum subtraction.}
\label{fig:q2233figs}
\end{figure}

%
The lack of Ly-$\alpha$ emitters in our fields, while surprising, does 
not contradict the space density of 
Ly-$\alpha$ emitters seen in other studies at similar redshifts 
(e.g. Stiavelli et al. 2001; Palunas et al. 2004). Both these studies 
found several LAEs, but were much larger in field of view and redshift 
depth than our study. The field of Stiavelli et al. was a non-absorber 
field. The field of Palunas et al. does show some sub-DLAs, but 
is known to be a filament with higher density than a typical field region.
 
Fig. 3 plots our APO FP SFR limits with our earlier NICMOS results at $z \sim 1.9$ 
(Kulkarni et al. 2000, 2001), and the results 
of other emission line searches in quasar absorbers 
(see references in Kulkarni et al. 2005b). All data points have been 
converted to a common cosmological model with $\Omega_{m} = 0.3$, 
$\Omega_{\Lambda} = 0.7$, and $H_{0} = 70$ km s$^{-1}$ Mpc$^{-1}$. 
The majority of the data in Fig.3  are 
for DLAs. We note, however, that many of the 
H-$\alpha$ and Ly-$\alpha$ candidates have not yet been spectroscopically 
confirmed. Furthermore, many of the H-$\alpha$ points are for objects at 
large angular separations from the quasar. Our APO FP limits 
are among the tightest existing constraints on SFRs in absorption-selected 
galaxies, but are clearly consistent with many other measurements. 
The solid curve shows the LD5 calculation of Bunker et al. for the predicted 
cross-section-weighted SFR
based on the closed-box Pei \& Fall (1995) model, for 
large proto-spirals (with space density equal to that of local spirals) 
for $q_{0}=0.5$ and $H_{0} = 70$ km s$^{-1}$ Mpc$^{-1}$. 
The dashed curve shows their H5 prediction for the hierarchical 
hypothesis (with a higher absorber space density in the past). 
Clearly, a large fraction of the observed SFR values fall below the 
prediction of the large-disk scenario, and several lie even below the 
hierarchical prediction. Even though the predicted SFRs for $\Omega_{m} = 0.3$ and  
$\Omega_{\Lambda} = 0.7$ would be somewhat lower, a large fraction of the absorbers would 
still have SFRs below the LD5 predictions. 

\begin{figure}
 \includegraphics[height=2.5in,width=3.5in,angle=0]{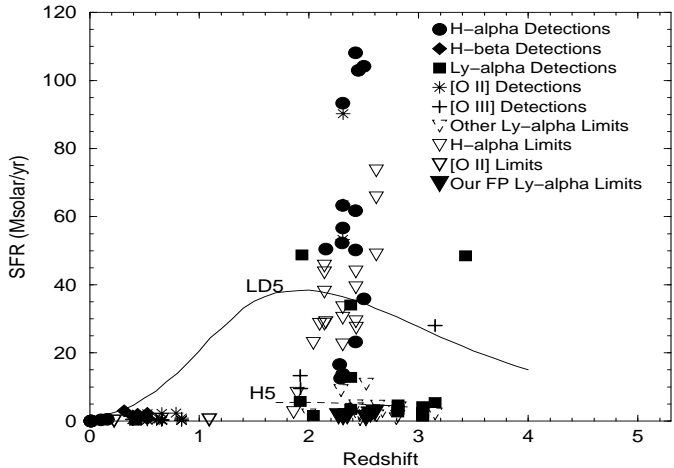}
  \caption{Measurements of star formation rates 
(in $M_{\odot}$ yr$^{-1}$) 
for confirmed or candidate objects in quasar 
absorber fields, from emission line searches. Data points are 
from our APO FP survey, our previous NICMOS 
imaging (Kulkarni et al. 2000, 2001), and other literature 
(Kulkarni et al. 2005b and references therein). The curves show 
the calculations of Bunker 
et al. (1999) for the predicted cross-section-weighted SFR in the large-disk 
and hierarchical scenarios.}\label{fig:vkulkarnifig3}
\end{figure}

\section{Conclusions}
Our results on both metallicities and SFRs appear to suggest that the history of 
metal production and star formation may have been quite different in DLAs 
than in the general galaxy population. Is this an effect of small number 
statistics? Or is this a selection effect caused by the fact that the 
more metal-rich and more vigorously star-forming DLAs are also likely to be 
dustier and may thus be underrepresented in flux-limited samples of quasars? 
Or are the star-forming regions in DLAs compact and at small angular 
separations from the quasars, so that they get lost in 
the quasar point spread function (PSF)? Clearly, it is necessary to expand the DLA metallicity 
samples at $z < 1.5$ to better understand the metallicity evolution. 
Such studies, together with 
high-resolution optical/IR imaging and spectroscopic confirmation of the 
absorbing 
galaxies will help to understand the 
overall role of DLAs in the big picture of galaxy evolution. 
\begin{acknowledgments}
We thank the organizers for holding this very stimulating 
conference. 
VPK, SMF, and JTL acknowledge partial 
support from the NASA/STScI grant GO-9441. 
VPK also acknowledges partial support from the U. S. 
National Science Foundation grant AST-0206197 and the Univ. of South Carolina 
Research Foundation. DEW acknowledges 
support from the NASA LTSA grant NAG5-11413. JWT acknowledges support of 
the NSF Physics Frontier Center, Joint Institute for Nuclear Astrophysics 
under grant PHY 02-16783 and DOE support under grant DE-FG 02-91ER 40606.
\end{acknowledgments}













\end{document}